\documentclass[final]{cvpr}

\usepackage{times}
\usepackage{epsfig}
\usepackage{graphicx}
\usepackage{amsmath}
\usepackage{amssymb}
\usepackage{subcaption}
\usepackage{adjustbox}
\usepackage{wrapfig,lipsum,booktabs} % for text side-by-side with the table
\usepackage[normalem]{ulem}
\usepackage[labelfont=bf]{caption}
\usepackage{array,multirow}
\usepackage{float}
\usepackage{booktabs}
\usepackage{multirow}
\usepackage[pagebackref=true,breaklinks=true,colorlinks,bookmarks=false]{hyperref}

\def\cvprPaperID{16} % *** Enter the CVPR Paper ID here
\def\confYear{CVPR 2021}

\begin{document}

%%%%%%%%% TITLE
\title{Unsupervised Detection of Cancerous Regions in Histology Imagery using Image-to-Image Translation}

\author{Dejan Štepec\textsuperscript{1,2}, Danijel Skočaj\textsuperscript{1}\\
\textsuperscript{1}University of Ljubljana, Faculty of Computer and Information Science\\
Večna pot 113, 1000 Ljubljana, Slovenia\\
\textsuperscript{2}XLAB d.o.o.\\
Pot za Brdom 100, 1000, Ljubljana, Slovenia\\
{\tt\small dejan.stepec@xlab.si}
}

%\author{Dejan Štepec\inst{1,2} \and Danijel Skočaj\inst{1}}
%\institute{University of Ljubljana, Faculty of Computer and Information Science\\
%Večna pot 113, 1000 Ljubljana, Slovenia
%\and
%XLAB d.o.o.\\
%Pot za Brdom 100, 1000, Ljubljana, Slovenia\\
%\email{dejan.stepec@xlab.si}}

\maketitle

%%%%%%%%% ABSTRACT
\begin{abstract}
   Detection of visual anomalies refers to the problem of finding patterns in different imaging data that do not conform to the expected visual appearance and is a widely studied problem in different domains. Due to the nature of anomaly occurrences and underlying generating processes, it is hard to characterize them and obtain labeled data. Obtaining labeled data is especially difficult in biomedical applications, where only trained domain experts can provide labels, which often come in large diversity and complexity. Recently presented approaches for unsupervised detection of visual anomalies approaches omit the need for labeled data and demonstrate promising results in domains, where anomalous samples significantly deviate from the normal appearance. Despite promising results, the performance of such approaches still lags behind supervised approaches and does not provide a one-fits-all solution. In this work, we present an image-to-image translation-based framework that significantly surpasses the performance of existing unsupervised methods and approaches the performance of supervised methods in a challenging domain of cancerous region detection in histology imagery.
\end{abstract}

%%%%%%%%% BODY TEXT
\section{Introduction}

The capability to detect anomalies in different data modalities has important applications in different domains, including medical imaging~\cite{anomaly_survey}. Detecting visual anomalies is a particularly challenging problem that has recently seen a significant rise of interest, due to the prevalence of deep-learning-based methods. Nevertheless, large part of this success can be attributed to the availability of large-scale labeled data, which is hard to obtain, as anomalies generally occur rarely, in different shapes and forms, and are thus extremely hard or even impossible to label.

Supervised deep-learning-based anomaly detection approaches have seen great success in different industrial and medical application domains~\cite{jama_dl, skocaj_anomaly}. The success of such methods is the most evident in the domains with well-known characterization (and possibly a finite set) of the anomalies and abundance of labeled data. Specific to the detection of visual anomalies, we usually also want to localize the actual anomalous region in the image. Obtaining such detailed labels to learn supervised models is a costly process and in many cases also impossible. There is an abundance of data available in the biomedical domain, but it is usually of much higher complexity and diversity. Usually only trained biomedical experts can annotate such data, preventing large-scale crowd annotation efforts.

Weakly-supervised approaches address such problems by requiring only image-level labels (e.g. disease present or not) and are able to detect and delineate anomalous regions solely from such weakly labeled data, without the need for detailed pixel or patch-level labels~\cite{weakly_nature}. On the contrary, few-shot approaches reduce the number of required labeled samples to the least possible amount~\cite{few-shot-polyp}, which can be further boosted with active learning, where the aim is to come up with the most informative and effective subset of samples for labeling~\cite{active_sa, active_ra}.

In an unsupervised setting, only normal appearance samples are available (e.g. healthy, defect-free), which are usually available in larger quantities and are easier to obtain. Deep generative methods, in a form of autoencoders (AE) or generative adversarial networks (GAN), have been recently applied to the problem of unsupervised detection of visual anomalies and have shown promising results in different industrial and medical application domains~\cite{f-anogan, anomaly_brain_scale, steganomaly, student_teacher}. Current approaches require normal appearance samples for training, in order to detect and segment deviations from that normal appearance, without the need for labeled data. They usually model normal appearance with low-resolution AE or GAN models and the overall performance still lags significantly behind supervised approaches.

In this work, we present a novel high-resolution image-to-image translation-based method for unsupervised detection of cancerous regions in histology imagery, that significantly surpasses the performance of existing GAN-based unsupervised approaches in that domain and also approaches the performance of the supervised counterparts.

%------------------------------------------------------------------------

\section{Related Work}

\subsection{Unsupervised Detection of Visual Anomalies} In an unsupervised setting, we only have access to normal samples (e.g. healthy, defect-free), which are used to model normal visual appearance. This is achieved by learning deep generative models, which results in the capability to generate realistic-looking artificial normal samples~\cite{stepec_anomaly, anogan}. An anomalous (i.e. out-of-distribution) sample is detected by comparing the original input (i.e. query sample) with its reconstruction, by thresholding on a domain-specific similarity metric. This is possible due to the learned manifold of normal appearance and its inability to reconstruct anomalous samples, resulting in higher visual dissimilarity~\cite{stepec_anomaly, anogan}.

Different approaches have been proposed for normal appearance modeling, as well as anomaly detection. Learning the normal visual appearance is based on autoencoders (AE)~\cite{anomaly_brain_scale}, generative adversarial networks (GAN)~\cite{anogan, f-anogan}, or combined hybrid models~\cite{ganomaly, skip_ganomaly} and was already investigated for the histology domain~\cite{stepec_anomaly}. Most of the approaches learn the space of the normal sample distribution $Z$, from which latent vectors $z \in Z$ are sampled, that generate the closest normal appearance, to the presented query image. Different solutions have been proposed for latent vector optimization, that are usually independent of the used normal appearance modeling method (i.e. AE, GAN).

Autoencoders (AE) represent an approach towards modeling normal appearance and have a major advantage with their ability to reconstruct images with low reconstruction errors, due to a supervised training objective. Unfortunately, they suffer from memorization and usually produce images that are blurry and of a much lower resolution, quality, and diversity, in comparison with GAN-based approaches~\cite{anomaly_brain}. Variational implementation (VAE) turns AE into a generative model, which enables modeling of data distribution and they are usually also combined with a GAN discriminator, in order to produce sharper images~\cite{anomaly_brain, ganomaly, skip_ganomaly}.

In comparison with autoencoders, GANs do not automatically yield the inverse mapping from the image to latent space, which is needed for closest-looking normal sample reconstruction and consequently anomaly detection. In AnoGAN~\cite{anogan} an iterative optimization approach was proposed to optimize the latent vector $z$ via backpropagation, using the residual and discrimination losses. In the f-AnoGAN method~\cite{f-anogan}, an autoencoder replaces the iterative optimization procedure, using the trainable encoder and the pre-trained generator (normal appearance modeling), as the decoder. In comparison with AnoGAN~\cite{anogan}, StyleGAN2~\cite{styleGAN2} enables high-resolution normal appearance modeling, while a similar iterative optimization procedure is used for latent space mapping.

\subsection{Image-to-Image Translation} The goal of image-to-image translation is to learn a mapping between an input image and an output image of different domains. In the supervised setting, paired corresponding images from different domains are available (e.g. grayscale-color of the same image) and conditional GANs~\cite{pix2pix} are used to learn the mapping. In the unsupervised setting, only two independent sets of images are available from source and target domains, with no paired examples (e.g. grayscale-color of different images). Cycle-consistency loss, presented with CycleGANs~\cite{cyclegan}, is a predominately used constraint for such inherently ill-posed problems. CycleGANs~\cite{cyclegan} enforce original image reconstruction, when mapping the source image to target domain and back, thus capturing special characteristics of the target domain and figuring it how to transfer it to the source domain, while preserving source domain image characteristics.

It was recently discovered, that CycleGANs are masters of steganography~\cite{steganography}, as it learns to "hide" information about the source image into images it generates, for almost perfect reconstruction. This intriguing property was exploited by SteGANomaly~\cite{steganomaly} for unsupervised anomaly detection. They used image-to-image translation to learn a mapping between the healthy brain MR images and a simulated distribution of healthy brain MR images with lower entropy. CycleGANs encode the source image information into the target image with a nearly imperceptible, high-frequency signal~\cite{steganography}, thus enabling the reconstruction of unseen anomalous samples during the inference. In SteGANomaly~\cite{steganomaly} they alleviate this problem by removing high frequency, low amplitude signal during training using Gaussian filtering in the target domain, before performing a complete cycle. The choice of intermediate distribution with lower entropy is very important, for the Gaussian filtering to effectively remove the hidden information, that is not relevant to the target domain. Together with the size of the Gaussian kernel, which needs to be manually fine-tuned, this represents a major limiting factor for general applicability.

\begin{figure*}[ht!]
    \centering
    \includegraphics[width=\linewidth]{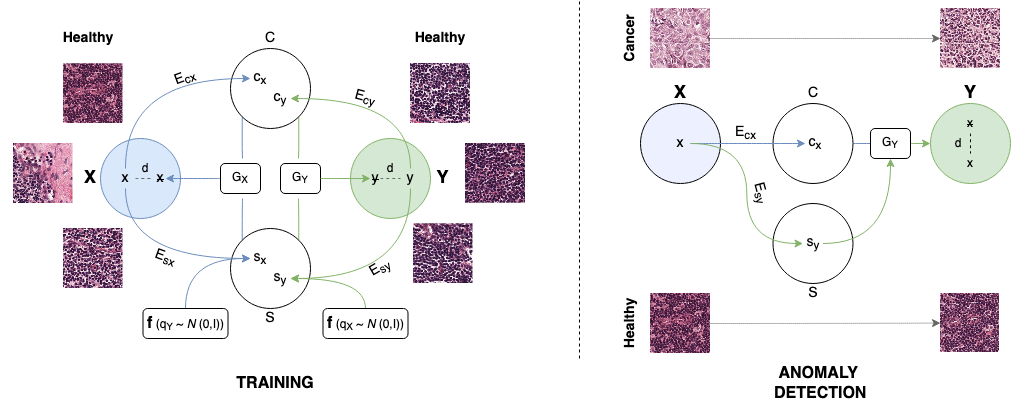}
    \caption{Our proposed unsupervised anomaly detection method, based on the image-to-image translation. We disentangle a latent space into shared content and style spaces, implemented via domain-specific (blue and green colors) encoders $E$ and decoders $G$. Anomaly detection is performed with an example-guided image translation.}
    \label{fig:proposed_method}
\end{figure*}

\subsection{Anomaly Detection in Medical Imagery} In this work, we particularly focus on a challenging task of cancerous region detection from gigapixel histology imagery, which has been already addressed in a supervised~\cite{jama_dl}, as well as in a weakly-supervised setting~\cite{weakly_nature}. There is a limited number of approaches in the literature that would approach that problem in an unsupervised fashion~\cite{histo_uad1, histo_uad2}. Extremely large histology imagery (patch-based processing) and the highly variable appearance of the different tissue regions represent a unique challenge for existing unsupervised approaches. Unsupervised approaches for the detection of visual anomalies are also applied to other biomedical imaging data, especially brain MRI data analysis, where brain lesion segmentation is the predominantly addressed problem~\cite{brainlesion, anomaly_brain, steganomaly}.

%------------------------------------------------------------------------

\begin{figure*}[htbp]
    \centering
    \begin{subfigure}{0.25\textwidth}
      \includegraphics[width=\linewidth]{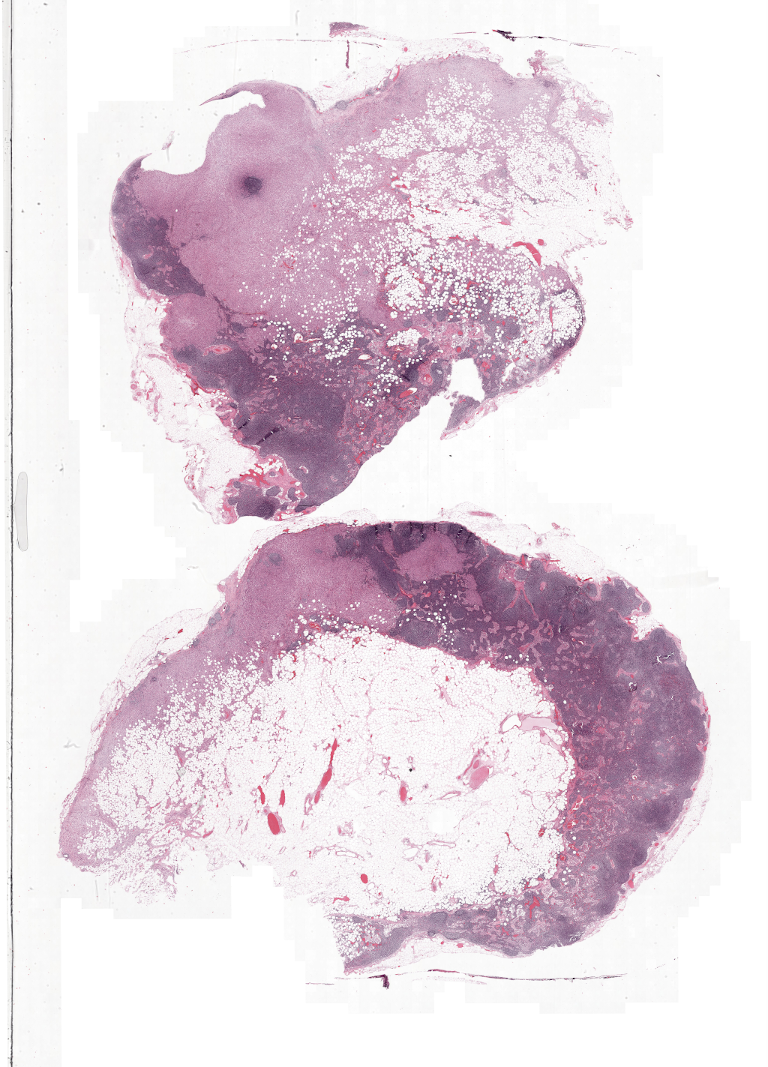}
      \caption{Original WSI}
      \label{fig:original_wsi}
    \end{subfigure}\hfil
        \begin{subfigure}{0.25\textwidth}
      \includegraphics[width=\linewidth]{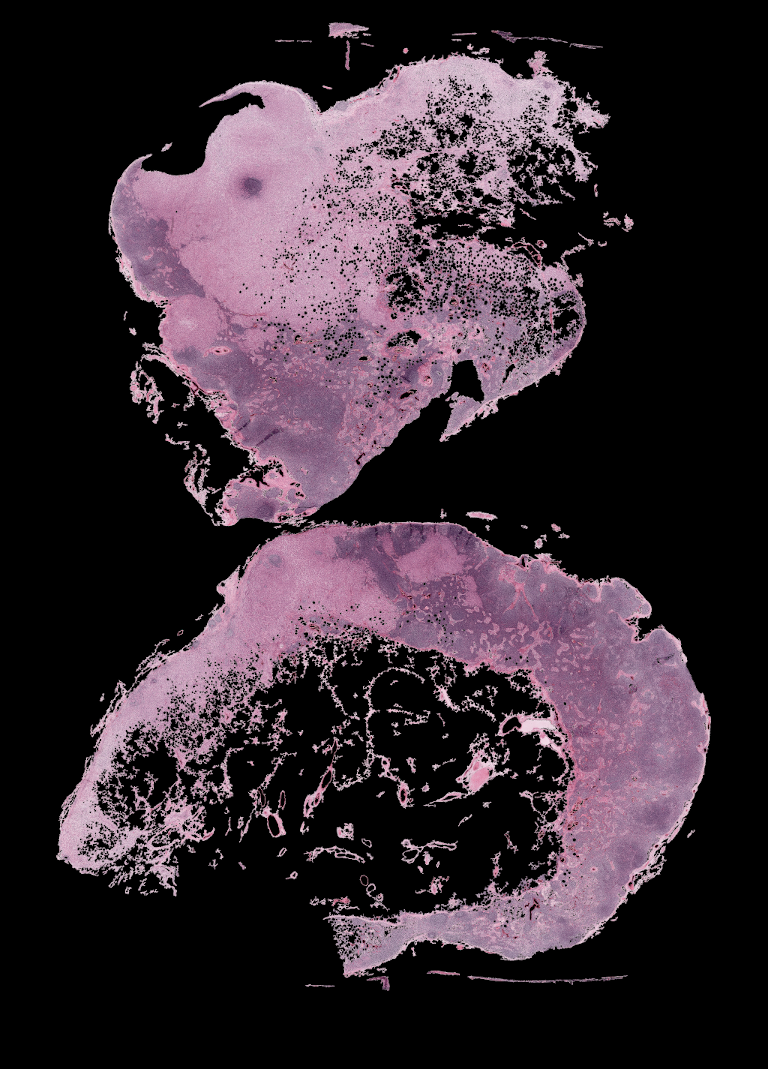}
      \caption{Filtered WSI}
      \label{fig:filtered_wsi}
    \end{subfigure}\hfil
    \begin{subfigure}{0.25\textwidth}
      \includegraphics[width=\linewidth]{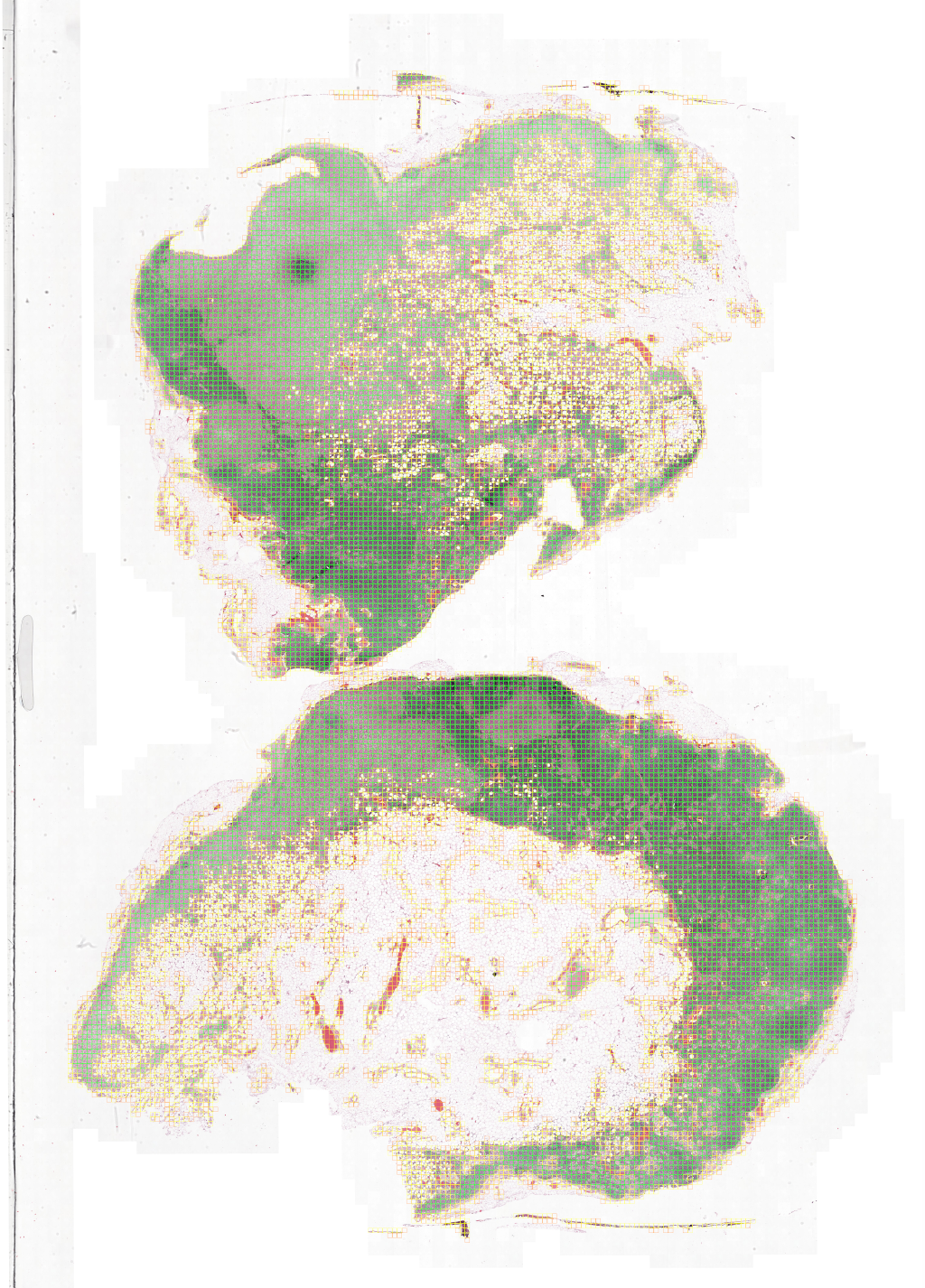}
      \caption{Tissue patches}
      \label{fig:tissue_patches}
    \end{subfigure}\hfil
    \begin{subfigure}{0.25\textwidth}
      \includegraphics[width=\linewidth]{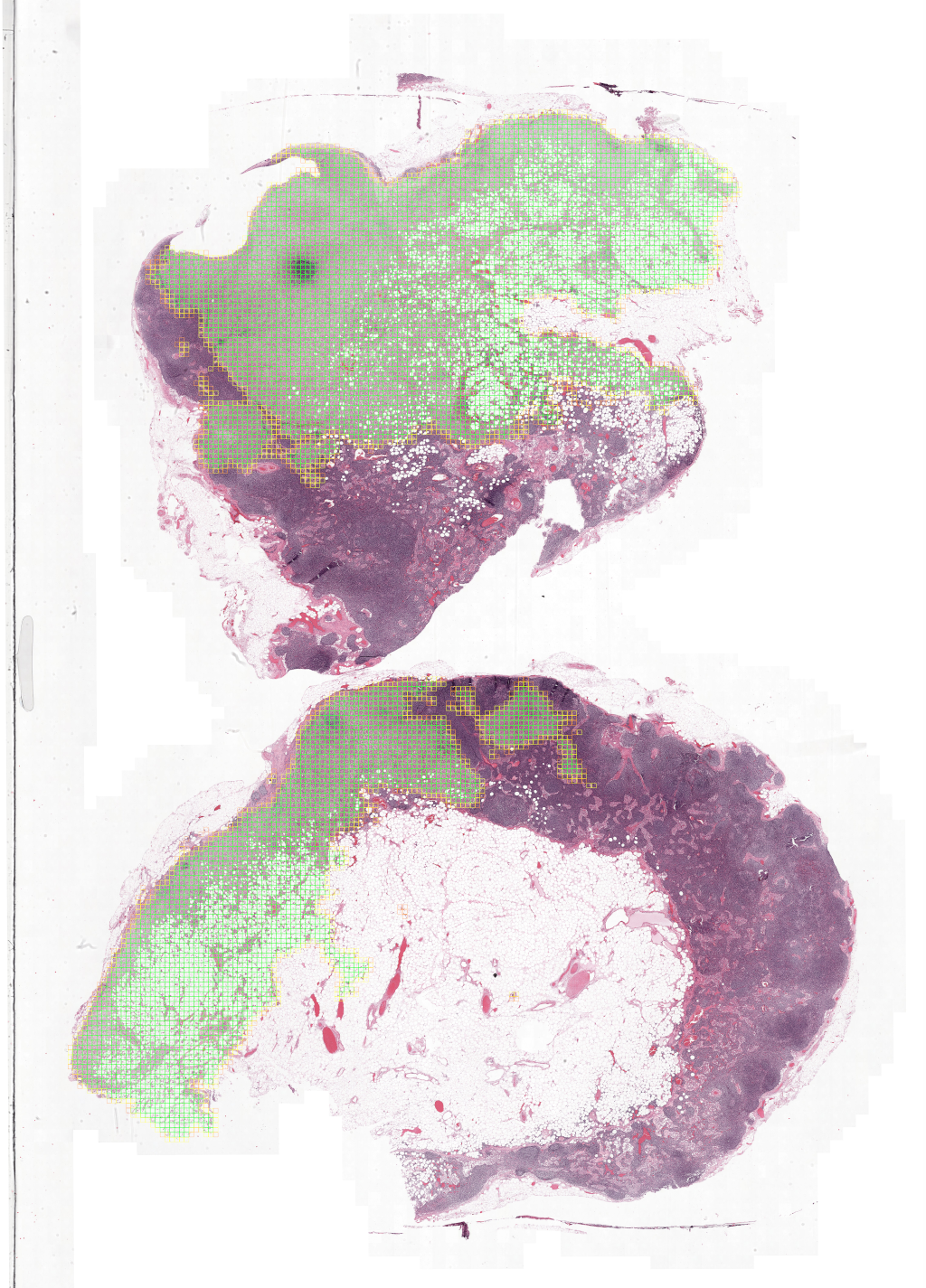}
      \caption{Cancer patches}
      \label{fig:cancer_patches}
    \end{subfigure}\hfil
    \caption{Preprocessing of the original WSI presented in a) consists of b) filtering tissue sections and c) extracting tissue patches, based on the tissue (green $\geq$ 90 \%, orange $\leq$ 10 \% and yellow in-between) and d) cancerous region coverage (green $\geq$ 90 \%, orange $\leq$ 30 \% and yellow in-between). Best viewed in a digital version with zoom.}
    \label{fig:preprocessing_wsi}
\end{figure*}

\section{Methodology}

We argue that image-to-image translation can be effective for unsupervised detection of visual anomalies and that this can be achieved with a direct mapping between unpaired sets of healthy cohorts, with an appropriate architecture, which successfully disentangles content, that needs to be preserved, apart from the style, which needs to change. CycleGAN~\cite{cyclegan} based methods perform this implicitly using the adversarial and cycle-consistency losses, which assumes a strong bijection between the two domains - resulting in steganography~\cite{steganography}. Inspired by the multimodal image-to-image translation methods~\cite{munit, drit++}, we propose an example guided image translation method (Figure~\ref{fig:proposed_method}), which in comparison with SteGANomaly~\cite{steganomaly} enables anomaly detection without cycle-reconstruction during the inference, specially crafted intermediate domain distribution, and Gaussian filtering. Similar to MUNIT~\cite{munit}, we assume that the latent space of images can be decomposed into content and style spaces. We also assume that images in both domains share a common content space $C$, as well as style space $S$. This differs from MUNIT~\cite{munit}, where style space is not shared, due to semantically different domains $X$ and $Y$. Similar to MUNIT~\cite{munit}, our translation model consists out of encoder $E_{ij}$ and decoder $G_{j}$ networks for each space $i \in \{C,S\}$ and domains $j \in \{X,Y\}$. Those subnetworks are used for autoencoding, as well as cross-domain translation, by interchanging encoders and decoders from different domains. Style latent codes $s_x$ and $s_y$ are randomly drawn for cross-domain translation and used as Adaptive Instance Normalization (AdaIN)~\cite{adain} parameters in residual blocks of decoders $G_x$ and $G_y$, additionally transformed by a multilayer perceptron (MLP) network $f$. For autoencoding, $E_{sx}$ and $E_{sy}$ encoders are used directly, to extract style codes. This randomness during cross-domain translation in training prevents the effect of memorization, largely present in autoencoder-based anomaly detection approaches. Different losses between inputs and its reconstructions (e.g. $x$ and \textit{\sout{x}}) are used to train autoencoders (L1 loss), cross-domain translations (adversarial loss), as well as reconstructions (cycle-consistency loss). The architecture and training objectives closely follow the implementation of MUNIT~\cite{munit}, with the exception of the selection of $X$ and $Y$ domains.

During anomaly detection (Figure~\ref{fig:proposed_method}), an input image $x$ is encoded with $E_{cx}$, to produce content vector $c_x$, which is then joined with the style code $s_y$, extracted from the original image $x$, with the style encoder $E_{sy}$ of the target domain $Y$. This presents an input to decoder $G_y$, which generates $y$. This is basically an example guided image translation, used also in MUNIT~\cite{munit} and DRIT++~\cite{drit++} methods. Content-style space decomposition is especially well suited for histopathological analysis due to different staining procedures, which causes the samples to significantly deviate in their visual appearance. Style-guided translation ensures that the closest looking normal appearance is found, taking into account also the staining appearance. We then measure an anomaly score using distance metric $d$ (e.g. perceptual LPIPS distance~\cite{lpips} or Structure Similarity Index (SSIM)~\cite{ssim_index}), between the original image $x$ and its reconstruction \textit{\sout{x}}.

%------------------------------------------------------------------------
\section{Experiments and Results}

\begin{figure*}[ht!]
    \centering
    \begin{subfigure}{0.25\textwidth}
      \includegraphics[width=\linewidth]{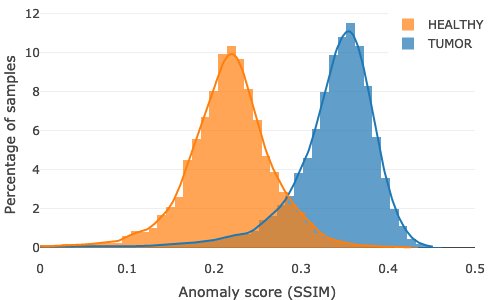}
      \caption{Proposed (SSIM)}
      \label{fig:dist_ours_ssim}
    \end{subfigure}\hfil
    \begin{subfigure}{0.25\textwidth}
      \includegraphics[width=\linewidth]{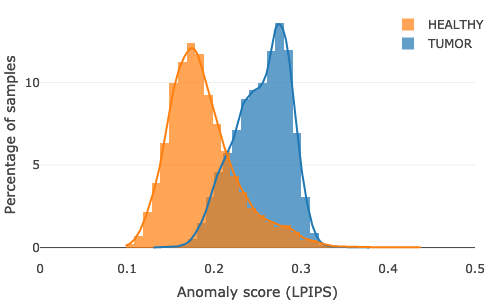}
      \caption{Proposed (LPIPS)}
      \label{fig:dist_ours_lpips}
    \end{subfigure}\hfil
    \begin{subfigure}{0.25\textwidth}
      \includegraphics[width=\linewidth]{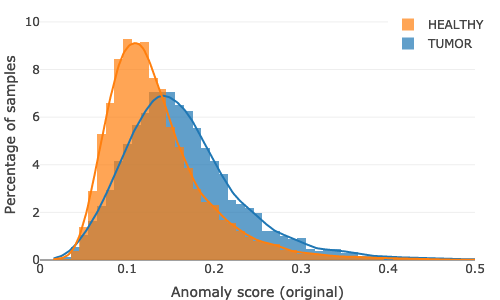}
      \caption{f-AnoGAN (original)}
      \label{fig:dist_fanogan}
    \end{subfigure}\hfil
    \begin{subfigure}{0.25\textwidth}
      \includegraphics[width=\linewidth]{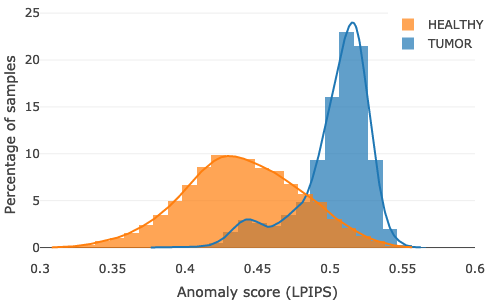}
      \caption{StyleGAN2 (LPIPS)}
      \label{fig:dist_stylegan2}
    \end{subfigure}\hfil
    \caption{Distribution of anomaly scores on healthy and cancerous histology imagery patches (a) for the proposed method (SSIM metric), (b) proposed method (LPIPS metric), (c)  f-AnoGAN (original metric), and (d) StyleGAN2 (LPIPS metric). Results for the proposed and StyleGAN2 methods are reported for $512^2$ patches, while $64^2$ patches are used for f-AnoGAN.}
   \label{fig:anomaly_dist}
\end{figure*}

\subsection{Histology Imagery Dataset} We evaluate the proposed anomaly detection pipeline on whole-slide histology images (WSI), which are used for diagnostic assessment of the spread of the cancer. This particular problem was already addressed in a supervised setting~\cite{jama_dl}, as a competition\footnote{https://camelyon16.grand-challenge.org/}, with provided clinical histology imagery and ground truth data. A training dataset with (n=110) and without (n=160) cancerous regions is provided, as well as a test set of 129 images (49 with and 80 without anomalies). Raw histology imagery, presented in Figure~\ref{fig:original_wsi}, is first preprocessed, in order to extract the tissue region (Figure~\ref{fig:filtered_wsi}). We used the approach from IBM\footnote{https://github.com/CODAIT/deep-histopath}, which utilizes a combination of morphological and color space filtering operations and was also used in prior work~\cite{stepec_anomaly} of synthesizing realistically looking histology samples. Patches of 512 x 512 are then extracted from the filtered image and filtered according to the tissue (Figure~\ref{fig:tissue_patches}) and cancer (Figure~\ref{fig:cancer_patches}) coverage. We only use patches with tissue and cancerous region coverage over 90 \% \linebreak(i.e. green patches).

%We address the aforementioned problems of anomaly detection pipeline on a challenging domain of digital pathology, where whole-slide histology images (WSI) are used for diagnostic assessment of the spread of cancer. This particular problem was already addressed in a %supervised setting~\cite{jama_dl}, as a competition\footnote{https://camelyon16.grand-challenge.org/}, with provided clinical histology imagery and ground truth data. A training dataset with (n=110) and without (n=160) cancerous regions is provided, as well as a test set of %129 images (49 with and 80 without anomalies). Raw histology imagery, presented in Figure~\ref{fig:original_wsi}, is first preprocessed, in order to extract the tissue region (Figure~\ref{fig:filtered_wsi}). We used the approach from %IBM\footnote{https://github.com/CODAIT/deep-histopath}, which utilizes a combination of morphological and color space filtering operations. Patches of 512 x 512 are then extracted from the filtered image and filtered according to the tissue (Figure~\ref{fig:tissue_patches}) %and cancer (Figure~\ref{fig:cancer_patches}) coverage. We only use patches with tissue and cancerous region coverage over 90 \% \linebreak(i.e. green patches).

We train the models on random 80,000 healthy tissue patches extracted from a training set of healthy and cancerous (cancer coverage=0\%) WSIs (n=270). The baseline supervised approach is trained on randomly extracted healthy (n=25,000) and cancerous patches (n=25,000). The methods are evaluated on healthy (n=7673) and cancerous (n=16,538) patches extracted from a cancerous test set of WSIs (n=49). We mix healthy training patches of both cohorts (i.e. healthy patches from cancerous WSIs) in order to demonstrate the robustness of the proposed approach against a small percentage of possibly contaminated healthy appearance data (e.g. non-labeled isolated tumor cells).

\subsection{Unsupervised Anomaly Detection} We compare the proposed method against GAN-based f-AnoGAN~\cite{f-anogan} and StyleGAN2~\cite{styleGAN2} methods. Both methods separately model normal appearance and perform latent space mapping for anomaly detection. The f-AnoGAN method models normal appearance using Wasserstein GANs (WGAN)~\cite{WGAN}, which is limited to a resolution of $64^2$ and uses an encoder-based fast latent space mapping approach. The StyleGAN2 method enables high-resolution image synthesis (up to $1024^2$) and also implements an iterative optimization procedure, based on Learned Perceptual Image Patch Similarity (LPIPS)~\cite{lpips} distance metric. We evaluate the performance of the proposed and StyleGAN2 methods on patches of $512^2$, while center-cropped $64^2$ patches are used for the f-AnoGAN method. Additionally, we compare the performance against the supervised DenseNet-121~\cite{densenet} baseline model, trained and evaluated on $512^2$ patches. We evaluate the proposed method using Structural Similarity Index Measure (SSIM)~\cite{ssim_index} and LPIPS reconstruction error metrics as an anomaly score. We use the same metrics as also as an alternative to original f-AnoGAN anomaly score implementation, as well as to measure StyleGAN2 reconstruction errors.

We first evaluate the methods by inspecting the distribution of anomaly scores across healthy and cancerous patches, as presented in Figure~\ref{fig:anomaly_dist}. We compare our proposed approach (Figures~\ref{fig:dist_ours_ssim} and~\ref{fig:dist_ours_lpips}) against f-AnoGAN (Figure~\ref{fig:dist_fanogan}) and StyleGAN2 (Figure~\ref{fig:dist_stylegan2}) methods and report significantly better distribution disentanglement.

\begin{figure}[ht!]
    \centering
    \includegraphics[width=1\linewidth]{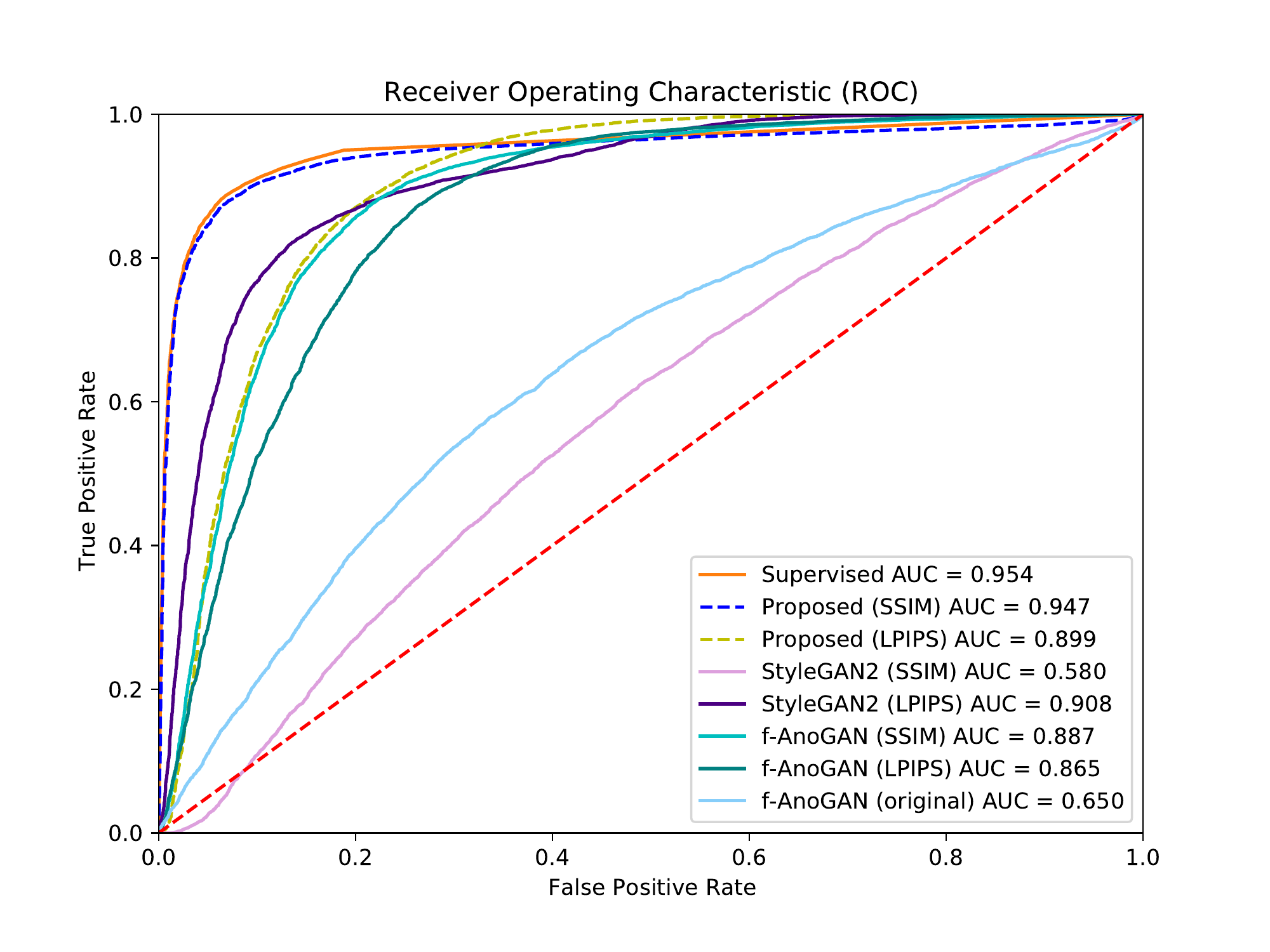}
    \caption{ROC curves for different methods and different reconstruction error metrics (i.e. anomaly scores).}
    \label{fig:roc_curves}
\end{figure}

Area under the ROC curve (AUC) scores are reported in Table~\ref{tbl:auc_results} for all the methods and different anomaly scores. Corresponding ROC curves are presented in Figure~\ref{fig:roc_curves}. We also report F1 and classification accuracy measures, calculated at the Youden index of the ROC curve. We notice that the performance of the proposed method approaches the performance of the supervised baseline in terms of both reconstruction error metrics (i.e. LPIPS and SSIM). The performance of the f-AnoGAN significantly improves using SSIM and LPIPS metrics, in comparison with the originally proposed anomaly score. This shows the importance of the selection of the appropriate reconstruction error metric. The styleGAN2 method shows good distribution disentanglement using the LPIPS distance metric, while the SSIM metric fails to capture any significant differences between the two different classes (i.e. healthy and anomalous). The proposed method demonstrates consistent performance across both anomaly score metrics, as well as different evaluation measures.

\setlength{\tabcolsep}{0.7\tabcolsep}
\begin{table}[ht!]
\centering
\caption{Performance statistics (F1, Classification Accuracy - CA) calculated at Youden index of Receiver Operating Characteristic (ROC) curve and the corresponding area under the ROC curve (AUC) and Average Precision (AP) scores summarizing ROC and Precision-Recall (PR) curves.}
\begin{tabular}{@{}lllll@{}}
\toprule
            & AUC   & AP & F1 & CA \\ \toprule
Supervised        & 0.954 &  0.974  &  0.925 &  0.901  \\ \midrule
Proposed (SSIM)          & \textbf{0.947} &  \textbf{0.976}  &  \textbf{0.920} &  \textbf{0.895}  \\
Proposed (LPIPS) & 0.900 &  0.914  &  0.886 &  0.847  \\ \midrule
StyleGAN2 (LPIPS) & 0.908 &  0.940  &  0.872 &  0.836  \\
StyleGAN2 (SSIM) & 0.580 &  0.711  &  0.674 &  0.588  \\ \midrule
f-AnoGAN (original) & 0.650 &  0.443  &  0.502 &  0.637 \\
f-AnoGAN (SSIM) & 0.887 &  0.916  &  0.886 &  0.846 \\
f-AnoGAN (LPIPS) & 0.865 &  0.902  &  0.875 &  0.830 \\ \bottomrule
\end{tabular}
\label{tbl:auc_results}
\end{table}

%\begin{table}[]
%\begin{tabular}{@{}l|ll|l@{}}
%\toprule
%                  & \multicolumn{2}{l|}{Anomaly score} & %\multirow{2}{*}{\begin{tabular}[c]{@{}l@{}}Supervised on \\ healed %images\end{tabular}} \\
%Method            & SSIM             & LPIPS           &        \\ \midrule
%Proposed          & 0.947            & 0.899           & \textbf{0.549}  \\        %                                 
%StyleGAN2         & 0.580            & 0.908           & 0.832  \\                 %                                
%f-AnoGAN ($64^2$) & 0.887            & 0.865           & ?      \\ \bottomrule
%\end{tabular}
%\end{table}

%%%%%%%%%%%%%%%%%%%%%%%%%%%%%%%%%%%%%%%%%%%%%%%%%

\begin{figure*}[ht!]
    \centering
    \includegraphics[width=\linewidth]{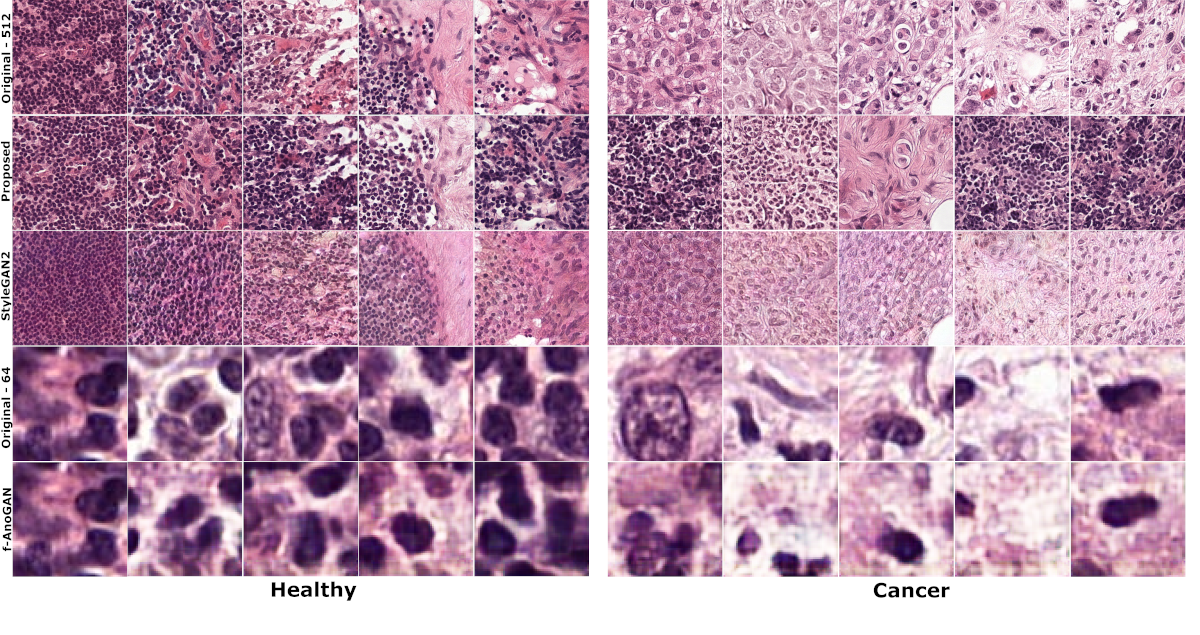}
    \caption{Example reconstructions of ground truth healthy and cancerous tissue samples with the proposed and StyleGAN2 methods for $512^2$ resolution and f-AnoGAN method for $64^2$. Best viewed in a digital version with zoom.}
    \label{fig:reconstructions}
\end{figure*}

In Figure~\ref{fig:reconstructions} we present example reconstructions of healthy and cancer tissue samples for all the methods. The proposed method reconstructs healthy samples much more accurately in comparison with the StyleGAN2 method. Some level of \textit{artificial healing} is visible on cancerous samples (i.e. visual appearance much more closely reassembles healthy samples). The f-AnoGAN method is only able to operate on $64^2$ resolution tissue samples, where similarly we notice better reconstructions of healthy appearances. The visual patterns in the StyleGAN2 method reconstructions demonstrate significantly lower variability in the visual appearance, especially in comparison with its demonstrated capability to synthesize realistically looking, highly variable, high-resolution histology tissue samples~\cite{stepec_anomaly}. In comparison, the proposed image-to-image translation-based enables high-resolution image synthesis, as well as example-based reconstruction, that can be effectively utilized for the detection of visual anomalies.

%------------------------------------------------------------------------
\section{Conclusion}

Detection of visual anomalies is an important process in many domains and recent advancements in deep generative-based methods have shown promising results towards applying them in an unsupervised fashion. This has sparked research in many domains, that did not benefit much from traditional supervised deep-learning-based approaches, due to difficulties in obtaining large quantities of labeled data. The medical image analysis domain is one such notable example, where the availability of imaging data in large quantities is usually not a problem, but the real challenge lies in the scarcity of human-expert annotations.

In this work, we presented an image-to-image translation-based unsupervised approach that significantly surpasses the performance of existing GAN-based unsupervised approaches for the detection of visual anomalies in histology imagery and also approaches the performance of supervised methods. The method is capable of closely reconstructing presented healthy histology tissue samples, while unable to reconstruct cancerous ones and is thus able to detect such samples with an appropriate visual distance measure.

The image-to-image translation-based framework offers a promising multi-task platform for a wide range of problems in the medical domain and can be now further extended with the capabilities for anomaly detection. Additional research is needed to investigate effectiveness in other biomedical modalities, as well as to exploit the benefits of using such a framework in a multi-task learning setting.

%------------------------------------------------------------------------

\section*{Acknowledgment}

This work was partially supported by the European Commission through the Horizon 2020 research and innovation program under grant 826121 (iPC) and by the Slovenian Research Agency (ARRS) project J2-9433 (DIVID).

{\small
\bibliographystyle{ieee_fullname}
\bibliography{cvpr}
}

\end{document}